\documentstyle[epsfig]{aipproc}

\def\Journal#1#2#3#4{{#1} {\bf #2}, #3 (#4)}

\def\NPB{{\em Nucl.~Phys.}~B}
\def\PLB{{\em Phys.~Lett.}~B}
\def\PRL{\em Phys.~Rev.~Lett.}
\def\PRD{{\em Phys.~Rev.}~D}

\def\be{\begin{equation}}
\def\ee{\end{equation}}
\def\bea{\begin{eqnarray}}
\def\eea{\end{eqnarray}}
\def\ba{\begin{array}}
\def\ea{\end{array}}
\def\nn{\nonumber}
\def\simge{\mathrel{%
   \rlap{\raise 0.511ex \hbox{$>$}}{\lower 0.511ex \hbox{$\sim$}}}}
\def\simle{\mathrel{
   \rlap{\raise 0.511ex \hbox{$<$}}{\lower 0.511ex \hbox{$\sim$}}}}
 
\def\slashchar#1{\setbox0=\hbox{$#1$}           
   \dimen0=\wd0                                 
   \setbox1=\hbox{/} \dimen1=\wd1               
   \ifdim\dimen0>\dimen1                        
      \rlap{\hbox to \dimen0{\hfil/\hfil}}      
      #1                                        
   \else                                        
      \rlap{\hbox to \dimen1{\hfil$#1$\hfil}}   
      /                                         
   \fi}                                         %
\def\ts{\thinspace}

\def\ra{\rightarrow}

\def\ol{\bar}

\def\CA{{\cal A}}
\def\CB{{\cal B}}
\def\CC{{\cal C}}
\def\CD{{\cal D}}

\def\CM{{\cal M}}

\def\CO{{\cal O}}

\def\ecm{\sqrt{s}}

\def\atc{\alpha_{TC}}

\def\atro{\alpha_{\rho_T}}
\def\Few{F_\pi}
\def\Ntc{N_{TC}}

\def\sutc{SU(\Ntc)}

\def\getc{g_{ETC}}

\def\thw{\theta_W}
\def\kslash{\raise.15ex\hbox{/}\kern-.57em k}
\def\LTC{\Lambda_{TC}}

\def\METC{M_{ETC}}

\def\condtc{\langle \ol T T \rangle_{TC}}
\def\condetc{\langle \ol T T \rangle_{ETC}}
\def\tro{\rho_{T}}

\def\troz{\rho_{T}^0}
\def\tropm{\rho_{T}^\pm}

\def\tom{\omega_T}
\def\tpi{\pi_T}
\def\tpipm{\pi_T^\pm}
\def\tpimp{\pi_T^\mp}
\def\tpip{\pi_T^+}
\def\tpim{\pi_T^-}
\def\tpiz{\pi_T^0}
\def\tpipr{\pi_T^{0 \ts\prime}}

\def\mm{\mu^+\mu^-}

\def\mev{{\rm MeV}}
\def\gev{{\rm GeV}}
\def\tev{{\rm TeV}}

\def\nb{{\rm nb}}
\def\pb{{\rm pb}}

\def\half{{\textstyle{ { 1\over { 2 } }}}}
\def\third{{\textstyle{ { 1\over { 3 } }}}}

\begin{document}

\title{Technicolor and the First~Muon~Collider
\thanks{Talk presented at the Workshop on Physics at the First Muon
  Collider and at the Front End of a Muon Collider}}

\author{Kenneth Lane}
\address{Department of Physics, Boston University, 590 Commonwealth Ave,
Boston, MA 02215}

\maketitle

\begin{abstract}

The motivations for studying dynamical scenarios of electroweak and
flavor symmetry breaking are reviewed and the latest ideas, especially
topcolor-assisted technicolor, are summarized. Technicolor's observable
low-energy signatures are discussed. The superb energy resolution of the
First Muon Collider may make it possible to resolve the extraordinarily
narrow technihadrons that occur in such models---$\tpiz$, $\troz$,
$\tom$---and produce them at very large rates compared to other colliders.

\end{abstract}

\section{Overview of Technicolor}\label{sec:tc}

Technicolor---the strong interaction of fermions and gauge bosons at the
scale $\LTC \sim 1\,\tev$---describes the breakdown of electroweak symmetry
to electromagnetism {\em without} elementary scalar bosons~\cite{tcref}.
In its simplest form, technicolor is a scaled-up version of QCD, with
massless technifermions whose chiral symmetry is spontaneously broken at
$\LTC$. If left and right-handed technifermions are assigned to weak
$SU(2)$ doublets and singlets, respectively, then $M_W = \cos\theta_W M_Z =
\half g \Few$, where $\Few = 246\,\gev$ is the {\em technipion} decay
constant,~\footnote{The only technipions in minimal technicolor are the
massless Goldstone bosons that become, via the Higgs mechanism, the
longitudinal components $W_L^\pm$ and $Z_L^0$ of the weak gauge bosons.} 
analogous to $f_\pi = 93\,\mev$ for the ordinary pion.

The principal signals in hadron and lepton collider experiments of
``classical'' technicolor were discussed long
ago~\cite{snowmass,ehlq}. In the minimal technicolor model, with
just one technifermion doublet, the only prominent collider signals are the
enhancements in longitudinally-polarized weak boson production.
These are the $s$-channel color-singlet technirho resonances near
1.5--2~TeV: $\troz \ra W_L^+W_L^-$ and $\tropm \ra W_L^\pm Z_L^0$. The
$\CO(\alpha^2)$ cross sections of these processes are quite small at such
masses. This and the difficulty of reconstructing weak-boson pairs with
reasonable efficiency make observing these enhancements a challenge.

Nonminimal technicolor models are much more accessible because they have a
rich spectrum of lower mass technirho vector mesons and technipion states
into which they may decay.~\footnote{The technipions of nonminimal
technicolor include the longitudinal weak bosons as well as additional
Goldstone bosons associated with spontaneous technifermion chiral symmetry
breaking. The latter must and do acquire mass---from the extended
technicolor interactions discussed below.} If there are $N_D$ doublets of
technifermions, all transforming according to the same complex
representation of the technicolor gauge group, there will be $4 N_D^2 - 1$
technipions whose decay constant is
\be\label{eq:ft}
F_T = {\Few \over{\sqrt{N_D}}} \ts .
\ee
Three of these are the longitudinal weak bosons; the remaining $4 N_D^2
- 4$ await discovery.

In the standard model and its extensions, the masses of quarks and leptons
are produced by their Yukawa couplings to the Higgs bosons---couplings of
arbitrary magnitude and phase that are put in by hand. This option is not
available in technicolor because there are no elementary scalars. Instead,
quark and lepton chiral symmetries must be broken explicitly {\it by gauge
interactions alone}. The most economical way to do this is to employ
extended technicolor, a gauge group containing flavor, color and
technicolor as subgroups~\cite{etcsd,etceekl,ichep}. Quarks, leptons and
technifermions are unified into a few large representations of ETC. The ETC
gauge symmetry is broken at high energy to technicolor $\otimes$ color.
Then quark and lepton hard masses arise from their coupling (with
strength $\getc$) to technifermions via ETC gauge bosons of generic mass
$\METC$:
\be\label{eq:qmass}
m_q(\METC) \simeq m_\ell(\METC)  \simeq {\getc^2 \over
{\METC^2}} \condetc \ts,
\ee
where $\condetc$ and $m_{q,\ell}(\METC)$ are the technifermion condensate
and quark and lepton masses renormalized at the scale $\METC$.

If technicolor is like QCD, with a running coupling $\atc$ rapidly becoming
small above $\LTC \sim 1\,\tev$, then $\condetc \simeq \condtc \simeq
\LTC^3$. To obtain quark masses of a few~GeV, $\METC/\getc \simle 30\,\tev$
is required. This is excluded: Extended technicolor boson exchanges also
generate four-quark interactions which, typically, include $|\Delta S| =
2$ and $|\Delta B| = 2$ operators. For these not to be in conflict with
$K^0$-$\ol K^0$ and $B_d^0$-$\ol B_d^0$ mixing parameters, $\METC/\getc$
must exceed several hundred TeV~\cite{etceekl}. This implies quark and
lepton masses no larger than a few MeV, and technipion masses no more than
a few~GeV---a phenomenological disaster.

Because of this conflict between constraints on flavor-changing neutral
currents and the magnitude of ETC-generated quark, lepton and technipion
masses, classical technicolor was superseded over a decade ago by
``walking'' technicolor~\cite{wtc}. Here, the strong technicolor coupling
$\atc$ runs very slowly---walks---for a large range of momenta, possibly
all the way up to the ETC scale of several hundred TeV. The slowly-running
coupling enhances $\condetc/\condtc$ by almost a factor of $\METC/\LTC$.
This, in turn, allows quark and lepton masses as large as a few~GeV and
$M_{\tpi} \simge 100\,\gev$ to be generated from ETC interactions at $\METC
= \CO(100\,\tev)$.

Walking technicolor requires a large number of technifermions in order that
$\atc$ runs slowly. These fermions may belong to many copies of the
fundamental representation of the technicolor gauge group, to a few higher
dimensional representations, or to both.~\footnote{The last possibility
inspired ``multiscale technicolor'' models containing both fundamental and
higher representations, and having an unusual phenomenology~\cite{multi}.
In multiscale models, there typically are two widely separated scales of
electroweak symmetry breaking, with the upper scale set by the weak decay
constant, $\Few = 246\,\gev$. Multiscale models in which the entire top
quark mass is generated by ETC interactions are excluded by such processes
as $b \ra s \gamma$~\cite{balaji}.}

In many respects, walking technicolor models are very different from QCD
with a few fundamental $SU(3)$ representations. One example of this is that
integrals of weak-current spectral functions and their moments converge
much more slowly than they do in QCD. Consequently, simple dominance of the
spectral integrals by a few resonances cannot be correct. This and other
calculational tools based on naive scaling from QCD and on large-$\Ntc$
arguments are suspect~\cite{glasgow}. Thus, it is not yet possible to predict
with confidence the influence of technicolor degrees of freedom on
precisely-measured electroweak quantities---the $S,T,U$ parameters to name
the most discussed example~\cite{pettests}.

The large mass of the top quark~\cite{toprefs} motivated another
major development in technicolor. Theorists have concluded that ETC models
cannot explain the top quark's large mass without running afoul of
experimental constraints from the $\rho$ parameter and the $Z \ra \ol b b$
decay rate~\cite{zbbth}. This state of affairs has led to the proposal of
``topcolor-assisted technicolor'' (TC2)~\cite{tctwohill}.

In TC2, as in top-condensate models of electroweak symmetry
breaking~\cite{topcondref}, almost all of the top quark mass arises from a
new strong ``topcolor'' interaction~\cite{topcref}. To maintain electroweak
symmetry between (left-handed) top and bottom quarks and yet not generate
$m_b \simeq m_t$, the topcolor gauge group under which $(t,b)$ transform is
usually taken to be a strongly-coupled $SU(3)\otimes U(1)$. The $U(1)$
provides the difference that causes only top quarks to condense. Then, in
order that topcolor interactions be natural---i.e., that their energy scale
not be far above $m_t$---without introducing large weak isospin violation,
it is necessary that electroweak symmetry breaking remain due mostly to
technicolor interactions~\cite{tctwohill}.

Early steps in the development of the TC2 scenario have been taken in two
recent papers\cite{tctwoklee}. The breaking of topcolor $SU(3)\otimes U(1)$
near the electroweak scale gives rise to a massive color octet of $V_8$
colorons and a color-singlet $Z'$. The $SU(3)$ may be broken by some of the
same technifermion condensates that break electroweak $SU(2) \otimes U(1)$,
so that the colorons (which are expected to be broad) have mass near
500~GeV. However, in order that the strong topcolor $U(1)$ interaction not
contaminate the ordinary $Z$ couplings to fermions, it and the weaker
$U(1)$ acting on light fermions must be broken down to their diagonal
subgroup, ordinary weak hypercharge, in the vicinity of 2~TeV. This
suggests that the $Z'$ mass is in the range 1--3~TeV, out of reach of
all but the highest energy colliders. As I discussed in my talk at the
FMC workshop, the $Z'$ is so heavy that it may require a multi-TeV Big
Muon Collider to find and study it. This subject deserves further study.

In TC2 models, ETC interactions are still needed to generate the light and
bottom quark masses, contribute a few~GeV to $m_t$,~\footnote{Massless
Goldstone ``top-pions'' arise from top-quark condensation. This ETC
contribution to $m_t$ is needed to give them a mass in the range of
150--250~GeV.} and give mass to the technipions. The scale of ETC
interactions still must be hundreds of~TeV to suppress flavor-changing
neutral currents and, so, the technicolor coupling still must walk.

Thus, even though the phenomenology of TC2 is still in its infancy, it is
expected to share general features with multiscale technicolor: many
technifermion doublets bound into many technihadron states, some at
relatively low masses, some carrying ordinary color and some not. The
lightest technihadrons may have masses in the range 100--300~GeV and
should be accessible at the Tevatron collider in Run~III if not
Run~II. All of them are easily produced and detected at the
LHC at moderate luminosities. If technihadrons exist, they will be
discovered at hadron colliders before the First Muon Collider (FMC) is
built. As we shall see, this is a good thing for the FMC: Several of the
lightest technihadrons are very narrow and can be produced in the
$s$-channel of $\mm$ annihilations. In the narrow-band FMC, it would be
exceedingly difficult to find them by a standard scan procedure without a
good idea of where to look.

\section{Technicolor at the FMC}\label{sec:tcfmc}

\subsection{Technihadron Decay Rates}

I assume that the technicolor gauge group is $\sutc$ and take $\Ntc =4$ in
calculations. Its gauge coupling must walk and I assume this is achieved by
a large number of isodoublets of technifermions transforming according to
the fundamental representation of $\sutc$. I consider the phenomenology of
only the lightest color-singlet technihadrons and assume that the
constraint from the $S$-parameter on their spectrum still allows the
lightest ones to be considered in isolation for a {\it limited} range of
$\ecm$, the $\mm$ center-of-mass energy, about their masses. These
technihadrons carry isospin $I = 1$ and~0 and consist of a single
isotriplet and isosinglet of vectors, $\troz$, $\tropm$ and $\tom$, and
pseudoscalars $\tpiz$, $\tpipm$, and $\tpipr$. The latter are in
addition to the longitudinal weak bosons, $W^\pm_L$ and $Z^0_L$---those
linear combinations of technipions that couple to the
electroweak gauge currents. I adopt TC2 as a guide for guessing
phenomenological generalities. In TC2 there is no need for large
technifermion isospin splitting associated with the top-bottom mass
difference. This implies that the lightest $\tro$ and $\tom$ are
approximately degenerate. The lightest charged and neutral technipions also
should have roughly the same mass, but there may be appreciable
$\tpiz$--$\tpipr$ mixing. If that happens, the lightest neutral technipions
are really $\ol U U$ and $\ol D D$ bound states. Finally, for purposes of
discussing signals at the FMC, we take the lightest technihadron masses to
be
\be\label{eq:tmasses}
M_{\tro} \cong M_{\tom} \sim 200\,\gev; \qquad
M_{\tpi} \sim 100\,\gev \ts.
\ee

The decays of technipions are induced mainly by ETC interactions which
couple them to quarks and leptons. These couplings are Higgs-like, and so
technipions are expected to decay into the heaviest fermion pairs allowed.
Because only a few GeV of the top-quark's mass is generated by ETC, there
is no great preference for $\tpi$ to decay to top quarks nor for top quarks
to decay into them. Furthermore, the isosinglet component of neutral
technipions may decay into a pair of gluons {\it if} its constituent
technifermions are colored. Thus, the predominant decay modes of the
light technipions are assumed to be
\be\label{eq:tpidecay}
\ba{l}
 \tpiz \ra \ol b b, \ts\ts \ol c c, \ts\ts \tau^+\tau^-  \\
 \tpipr \ra gg, \ts\ts \ol b b, \ts\ts \ol c c, \ts\ts \tau^+\tau^- \\
 \tpip \ra c \ol b, \ts \ts c \ol s, \ts\ts \tau^+\nu_\tau \ts. \\
\ea
\ee
To estimate branching ratios we use the following decay rates (for
later use in the technihadron production cross sections, we quote 
the energy-dependent width~\cite{ehlq,ellis}):~\footnote{The amplitude
is taken to be $\CM(\tpi \ra \ol f'(p_1) f(p_2)) = C_f (m_f + m_{f'})/F_T
\ts \ol u(p_2) \gamma_5 v(p_1)$.}
\bea\label{eq:tpiwidths}
 \Gamma(\tpi \ra \ol f' f) &=& {1 \over {16\pi F^2_T}}
 \ts N_f \ts p_f \ts C^2_f (m_f + m_{f'})^2 \nn \\ \nn \\
 \Gamma(\tpipr \ra gg) &=& {1 \over {128 \pi^3 F^2_T}} 
 \ts \alpha^2_S \ts C_{\tpi} \ts \Ntc^2 \ts s^{{3\over{2}}} \ts .
\eea
Here, $C_f$ is an ETC-model dependent factor of order one
{\it except} that TC2 suggests $\vert C_t\vert \simle m_b/m_t$; $N_f$ is
the number of colors of fermion~$f$; $p_f$ is the fermion momentum;
$\alpha_S$ is the QCD coupling evaluated at $M_{\tpi}$; and $C_{\tpi}$
is a Clebsch of order one. For $M_{\tpi} = 110\,\gev$, $F_T = \Few/3 =
82\,\gev$, $m_b = 4.2\,\gev$, $\Ntc = 4$, $\alpha_S = 0.1$, $C_b = 1$
for $\tpiz$ and $\tpipr$, and $C_{\tpi} = 4/3$:
\be\label{eq:tpiwidthvals}
\ba{l}
\Gamma(\tpiz \ra \ol b b) = \Gamma(\tpipr \ra \ol b b) =  35\,\mev\\ \\
\Gamma(\tpipr \ra gg) =  10\,\mev \ts.\\
\ea
\ee

If technicolor were like QCD, we would expect the main decay modes of the
lightest technivector mesons to be $\troz \ra \tpip\tpim$ and $\tom \ra
\tpip\tpim\tpiz$ with the technihadrons all composed of the same
technifermions. However, the large ratio $\condetc/\condtc$ occurring in
walking technicolor significantly enhances technipion masses compared to
technivector masses. Thus, $\tro \ra \tpi\tpi$ decay channels may well be
closed. If this happens, then $\troz$ decays to $W^+_L W^-_L$ or $W^\pm_L
\tpimp$ and $\tom$ to $\gamma \tpiz$ or $Z^0
\tpiz$.~\cite{multi,tpitev,elwtev,ichep}

We parameterize this for $\tro$ decays by adopting a simple model of two
isotriplets of technipions which are mixtures of $W_L^\pm$, $Z_L^0$ and
mass-eigenstate technipions $\tpipm$, $\tpiz$. The lighter isotriplet
$\tro$ is assumed to decay dominantly into pairs of the mixed state of
isotriplets $\vert\Pi_T\rangle = \sin\chi \ts \vert W_L\rangle + \cos\chi
\ts \vert\tpi\rangle$, where
\be\label{eq:sinchi}
\sin\chi = F_T/\Few \ts.
\ee
Then, the energy-dependent decay rate for $\troz \ra \pi_A^+ \pi_B^-$
(where $\pi_{A,B}$ may be $W_L$, $Z_L$, or $\tpi$) is given by
\be\label{eq:trhopipi}
\Gamma(\troz \ra \pi_A^+ \pi_B^-) = {2 \atro \CC^2_{AB}\over{3}} \ts
{\ts\ts p_{AB}^3\over {s}} \ts,
\ee
where $p_{AB}$ is the technipion momentum and $\atro$ is obtained by {\it
naive} scaling from the QCD coupling for  $\rho \ra \pi\pi$:
\be\label{eq:alpharho}
\atro = 2.91 \left({3\over{\Ntc}}\right)\ts.
\ee
The parameter $\CC^2_{AB}$ is given by
\be\label{eq:ccab}
\ba{ll}
\CC^2_{AB} =  &\left\{\ba{ll} \sin^4\chi &\mbox{for $W_L^+ W_L^-$}  \\
2\sin^2\chi \ts \cos^2\chi &\mbox{for $W_L^+ \tpim + W_L^- \tpip$} \\
\cos^4\chi &\mbox{for $\tpip \tpim$} 
\ea \right.\\
\ea
\ee
Note that the $\tro$ can be {\it very} narrow. For $\ecm = M_{\tro} =
210\,\gev$, $M_{\tpi} = 110\,\gev$, and $\sin\chi = \third$, we have
$\sum_{AB} \ts \Gamma(\troz\ra\pi^+_A\pi^-_B) = 680\,\mev$, 80\% of which
is $W_L^\pm \tpimp$.

We shall also need the decay rates of the $\tro$ to fermion-antifermion
states. The energy-dependent widths are
\be\label{eq:trhoff}
\Gamma(\troz \ra \ol f_i f_i) = {N_f\ts \alpha^2 \over
{3\atro}} \ts {p_i \ts (s + 2m^2_i) \over {s}}\ts A^0_i(s) \ts.
\ee
Here, $\alpha$ is the fine-structure constant, $p_i$ is the momentum and $m_i$
the mass of fermion $f_i$, and the factors $A^0_i$ are given by
\bea\label{eq:afactors}
A_i^0(s) &=& \vert \CA_{iL}(s) \vert^2
+ \vert \CA_{iR}(s) \vert^2 \ts, \nn \\
\CA_{i\lambda}(s) &=& Q_i + {2 \cos 2\thw \over {\sin^2 2\thw}} \ts
\zeta_{i \lambda} \left({s \over {s - M_Z^2 + i\ecm \ts
\Gamma_Z}}\right)\ts, \\
\zeta_{i L} &=& T_{3i} - Q_i \sin^2\thw, \qquad \zeta_{i R} = - Q_i
\sin^2\thw \ts. \nn
\eea
For $M_{\tro} = 210\,\gev$ and other parameters as above, the $\ol f f$
partial decay widths are:
\be\label{eq:trhoffvals}
\ba{ll}
\Gamma(\troz \ra \ol u_i u_i) = 5.8\,\mev \ts, \qquad
&\Gamma(\troz \ra \ol d_i d_i) = 4.1\,\mev \\
\Gamma(\troz \ra \ol \nu_i \nu_i) = 0.9\,\mev \ts, \qquad
&\Gamma(\troz \ra  \ell_i^+ \ell_i^-) = 2.6\,\mev \ts.
\ea
\ee

For the $\tom$, phase space considerations suggest we consider only its
$\gamma \tpiz$ and fermionic decay modes. The energy dependent widths are:
\bea\label{eq:tomegawidth}
\Gamma(\tom \ra \gamma \tpiz) &=& {\alpha p^3 \over {3 M_T^2}} \ts,
\nn \\ \nn \\
\Gamma(\tom \ra \ol f_i f_i) &=& {N_f \ts \alpha^2 \over
{3 \atro}} \ts {p_i\ts (s + 2m^2_i) \over {s}}\ts B^0_i(s) \ts.
\eea
The mass parameter $M_T$ in the $\tom \ra \gamma\tpiz$ rate is unknown
{\it a priori}; naive scaling from the QCD decay, $\omega \ra \gamma
\pi^0$, suggests it is several 100~GeV. The factor $B_i^0$ is given by
\bea\label{eq:bfactors}
B_i^0(s) &=& \vert \CB_{iL}(s) \vert^2 + \vert \CB_{iR}(s) \vert^2 \ts, 
\nn \\
\CB_{i\lambda}(s) &=& \left[Q_i - {4 \sin^2\thw \over {\sin^2 2\thw}} \ts
\zeta_{i\lambda} \left({s \over {s - M_Z^2 + i\ecm\ts \Gamma_Z}}
\right)\right] \ts (Q_U + Q_D) \ts.
\eea
Here, $Q_U$ and $Q_D = Q_U - 1$ are the electric charges of the
$\tom$'s constituent technifermions. For $M_{\tom} = 210\,\gev$ and
$M_{\tpi} = 110\,\gev$, and choosing $M_T = 100\,\gev$ and $Q_U = Q_D + 1
= {4\over {3}}$, the $\tom$ partial widths are:
\be\label{eq:tomegadkvals}
\ba{ll}
\Gamma(\tom \ra \gamma \tpiz) = 115\,\mev \\
\Gamma(\tom \ra \ol u_i u_i) = 6.8\,\mev \ts, \qquad
&\Gamma(\tom \ra \ol d_i d_i) = 2.6\,\mev \\
\Gamma(\tom \ra \ol \nu_i \nu_i) = 1.7\,\mev \ts, \qquad
&\Gamma(\tom \ra  \ell_i^+ \ell_i^-) = 5.9\,\mev \ts.
\ea
\ee

The beam momentum spread of the First Muon Collider has been quoted to
be as narrow as $\sigma_p/p = 3\times 10^{-5}$ at $\ecm = 100\,\gev$ and
$10^{-3}$ at $\ecm = 200\,\gev$. These correspond to beam energy spreads of
$\sigma_E = 5\,\mev$ at 100~GeV and 300~MeV at 200~GeV. The resolution at
100~GeV is less than the expected $\tpiz$, $\tpipr$ widths. At 200~GeV
it is sufficient to resolve the $\troz$, but not the $\tom$, for the
parameters we used. It is very desirable, therefore, that the 200~GeV FMC's
energy spread be about factor of~10 smaller. Since each of these
technihadrons can be produced as an $s$-channel resonance in $\mm$
annihilation, it would then be possible to sit on the peak at $\ecm =
M$. As we see next, the peak cross sections are enormous, 2--3 orders of
magnitude larger than can be achieved at a hadron collider and even at a
linear $e^+e^-$ collider because of the latter's inherent beam energy
spread.

\subsection{Technihadron Production Rates}

Like the standard Higgs boson, neutral technipions are expected to couple
to $\mm$ with a strength proportional to $m_\mu$. Compared to the Higgs,
however, this coupling is enhanced by a factor of $\Few/F_T = 1/\sin\chi$.
This makes the resolution of the FMC well-matched to the $\tpiz$ width.
Thus, the FMC is a technipion factory, overwhelming the rate at any other
collider. Once a neutral technipion has been found in $\tro$ or $\tom$
decays at a hadron collider, it should be relatively easy in the FMC to
locate the precise position of the resonance and sit on it. The cross
sections for $\ol f f$ and $gg$ production are isotropic; near the
resonance, they are given by
\bea\label{eq:tpirates}
{d\sigma(\mm \ra \tpiz \ts {\rm or} \ts \tpipr \ra \ol f f) \over{dz}} &=&
{N_f \over {2\pi}} \ts \left({C_\mu C_f m_\mu m_f \over
  {F_T^2}}\right)^2 \ts
{s \over{(s - M_{\tpi}^2)^2 + s \ts \Gamma_{\tpi}^2}} \ts,
\nn \\ \\
{d\sigma(\mm \ra \tpipr \ra gg) \over{dz}} &=&
{C_{\tpi} \over {32\pi^3}} \ts \left({C_\mu m_\mu\alpha_S\Ntc\over{F_T^2}}
\right)^2 \ts {s^2 \over{(s - M_{\tpi}^2)^2 + s \ts \Gamma_{\tpi}^2}}
\ts. \nn
\eea
Here, $z = \cos\theta$ where $\theta$ is the center-of-mass production
angle.

\begin{figure} 
\vspace{10pt}
\centerline{\epsfig{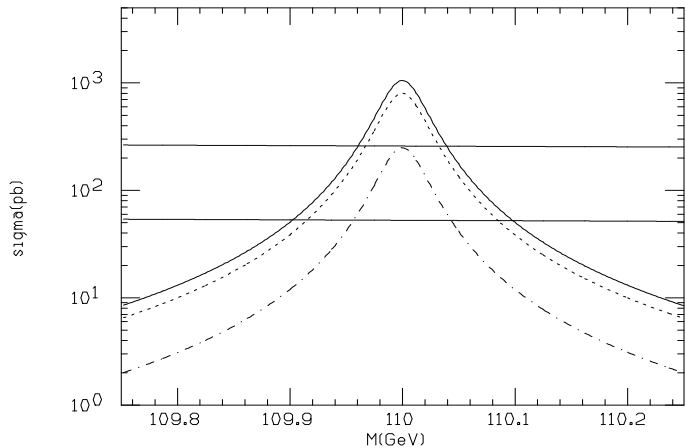}}
\vspace{10pt}
\caption{Theoretical (unsmeared) cross sections for $\mm \ra \tpipr \ra
\ol b b$ (dashed), $gg$ (dot-dashed) and total (solid) for $M_{\pi_T}=
110\,\gev$ and other parameters defined in the text. The solid
horizontal lines are the backgrounds from $\gamma, \ts Z^0 \ra \ol b b$
(lower) and $Z^0 \ra \ol q q$ (upper). Note the energy scale.}
\label{fig1}
\end{figure}

The $\tpipr$ production cross sections and the $Z^0$ backgrounds
are shown in Fig.~1 for $M_{\tpi} = 110\,\gev$ and other parameters as
above ($C_\mu = C_f = 1$, $C_{\tpi} = 4/3$, $F_T = 82\,\gev$, $\alpha_S
= 0.1$, $\Ntc = 4$). The peak signal rates approach $1\,\nb$.
The $\ol b b$ dijet rates are much larger than the
$Z^0 \ra \ol b b$ backgrounds, while the $gg$ rate is comparable to $Z^0
\ra \ol q q$. Details of these and the other calculations in this
section, including the effects of the finite beam energy resolution,
will appear in Ref.~\cite{elwfmc}. See Ref.~\cite{bdcth} for another 
example of neutral scalars that may be produced in $\mm$ annihilation.

The cross sections for technipion production via the decay of
technirho and techniomega $s$-channel resonances are calculated using
vector meson ($\gamma$, $Z^0$) dominance~\cite{ehlq,multi,tpitev,elwtev}.
They are given by:
\bea\label{eq:tvectorrates}
{d\sigma(\mm \ra \troz \ra \pi_A  \pi_B ) \over{dz}} &=& {\pi \alpha^2
p^3_{AB} \over {s^{1\over{2}}}} \ts {A^0_\mu(s) \ts \CC^2_{AB} \ts (1-z^2)
\over {(s - M_{\tro}^2)^2 + s \ts \Gamma_{\tro}^2}} \ts,
\nn \\ \nn \\
{d\sigma(\mm \ra \tom \ra \gamma \tpiz) \over{dz}} &=& {\pi \alpha^3
s^{1\over{2}} p^3 \over {3\atro M_T^2}} \ts {B^0_\mu(s) \ts (1+z^2)
\over {(s - M_{\tom}^2)^2 + s \ts \Gamma_{\tom}^2}} \ts,
\eea
where $A^0_\mu$ and $B^0_\mu$ were defined in Eqs.~\ref{eq:afactors} and
\ref{eq:bfactors}, respectively. For $M_{\tro} = M_{\tom} = 210\,\gev$,
$M_{\tpi} = 110\,\gev$, and other parameters as above, the total peak cross
sections are~\cite{elwfmc}:
\be\label{eq:tvectorratevals}
\ba{ll}
&\sum_{AB} \ts \sigma(\mm \ra \troz \ra \pi_A \pi_B)  = 1.1\,\nb \\ \\
&\sigma(\mm \ra \tom \ra \gamma \tpiz)  = 8.9\,\nb \ts.
\ea
\ee
The technirho rate is 20\% $W^+W^-$ and 80\% $W^\pm \tpimp$.

\begin{figure} 
\vspace{10pt}
\centerline{\epsfig{file=muon_ffbar_fmc.epsi,height=3.25in,width=2.5in}}
\vspace{10pt}
\caption{Theoretical (unsmeared) cross sections for $\mm \ra \troz, \tom \ra
e^+e^-$ for input masses $M_{\tro} = 210\,\gev$ and $M_{\tom} =
212.5\,\gev$ and other parameters as defined in the text.}
\label{fig2}
\end{figure}

Finally, it is reasonable to expect a small nonzero isospin splitting
between $\troz$ and $\tom$. This would appear as a dramatic interference in
the $\mm \ra \ol f f$ cross section {\it provided} the FMC energy
resolution is good enough in the $\tro$--$\tom$ region. The cross section
is most accurately calculated by using the full
$\gamma$--$Z^0$--$\tro$--$\tom$ propagator matrix, $\Delta(s)$. With
$\CM^2_V = M^2_V - i \ecm \ts \Gamma_V(s)$ for $V = Z^0,\tro,\tom$, this
matrix is the inverse of
\be\label{eq:vprop}
\Delta^{-1}(s) =\left(\ba{cccc}
s & 0 & -s f_{\gamma\tro} & -s f_{\gamma\tom} \\
0 & s - \CM^2_Z  & -s f_{Z\tro} & -s f_{Z\tom} \\
-s f_{\gamma\tro}  & -s f_{Z\tro}  & s - \CM^2_{\tro} & 0 \\
-s f_{\gamma\tom}  & -s f_{Z\tom}  & 0 & s - \CM^2_{\tom} 
\ea\right) \ts.
\ee
Here,
\bea\label{eq:ffactors}
f_{\gamma\tro} &=& \sqrt{{\alpha\over{\atro}}} \ts,
\qquad\qquad\qquad f_{\gamma\tom} = \sqrt{{\alpha\over{\atro}}} \ts (Q_U
+ Q_D)
\nn \\ \nn \\
f_{Z\tro} &=& \sqrt{{\alpha\over{\atro}}} \ts {\cos
  2\thw\over{\sin 2\thw}}
\ts, 
\qquad\ts  f_{Z\tom} = - \sqrt{{\alpha\over{\atro}}} \ts
{\sin^2\thw\over{\sin 2\thw}} \ts (Q_U + Q_D) \ts.
\eea
Then, the cross section is given in terms of matrix elements of $\Delta$ by
\bea\label{eq:mmffrate}
{d\sigma(\mm \ra \troz,\ts \tom \ra \ol f_i f_i) \over{dz}} &=&
{N_f \pi \alpha^2\over{8s}} \biggl\{
\left(\vert\CD_{iLL}\vert^2 + \vert\CD_{iRR}\vert^2\right)(1+z)^2 \nn \\
&& +\left(\vert\CD_{iLR}\vert^2 + \vert\CD_{iRL}\vert^2\right)(1-z)^2
\biggr\} \ts,
\eea
where
\bea\label{eq:dfactors}
\CD_{i\lambda\lambda'}(s) &=& s\biggl[Q_i Q_\mu \ts \Delta_{\gamma\gamma}(s)
 + {4\over{\sin^2 2\thw}} \ts \zeta_{i \lambda}
\ts \zeta_{\mu \lambda'} \ts \Delta_{ZZ}(s) \nn \\
&& + {2\over{\sin 2\thw}} \ts \biggl(\zeta_{i \lambda} Q_\mu
\Delta_{Z\gamma}(s) + Q_i \zeta_{\mu \lambda'} \Delta_{\gamma Z}(s)\biggr)
\biggr]
\ts.
\eea

Figure~2 shows the theoretical $\troz$--$\tom$ interference effect in
$\mm \ra e^+e^-$ for input masses $M_{\tro} = 210\,\gev$ and $M_{\tom} =
212.5\,\gev$ and other parameters as above. The propagator $\Delta$
shifts the nominal positions of the resonance peaks by
$\CO(\alpha/\atro)$. The theoretical peak cross sections are
$5.0\,\pb$ at $210.7\,\gev$ and $320\,\pb$ at $214.0\,\gev$. This
demonstrates the importance of precise resolution in the $200\,\gev$
FMC.

\section{Conclusions}\label{sec:close}

Modern technicolor models predict narrow neutral technihadrons, $\tpi$,
$\tro$ and $\tom$. These states would appear as spectacular resonances in a
$\mm$ collider with $\ecm = 100$--$200\,\gev$ and energy  resolution
$\sigma_E/E \simle 10^{-4}$. This is a very strong physics motivation for
building the First Muon Collider.

I thank other members of the First Muon Collider Workshop Strong Dynamics
Subgroup for valuable interactions, especially Paul Mackenzie and Chris
Hill for stressing the importance of a narrow $\tpiz$ at a muon collider,
and Pushpa Bhat, Estia Eichten and John Womersley for considerable
guidance. I am indebted to Torbjorn Sjostrand for first pointing out to me
the likely importance of the $\troz$ and $\tom$ decays to fermions. This 
led to our consideration of $\troz$--$\tom$ interference, a phenomenon 
which may actually be observable for the first time in a muon collider.



\begin{thebibliography}{99}
%
\bibitem{tcref}S.~Weinberg, \Journal{\PRD}{19}{1277}{1979}; L.~Susskind,
\Journal{\PRD}{20}{2619}{1979}.
%
\bibitem{snowmass}K.~Lane, {\it The Scalar Sector of the Electroweak
Interactions}, Proceedings  of the 1982 DPF Summer Study on Elementary
Particle Physics and Future  Facilities, edited by R.~Donaldson,
R.~Gustafson and F.~Paige (Fermilab 1983), p.~222.
%
\bibitem{ehlq}E.~Eichten, I.~Hinchliffe, K.~Lane and C.~Quigg, {\em
Rev.~Mod.~Phys.}~{\bf 56}, 579 (1984); \Journal{\PRD}{34}{1547}{1986}.
%
\bibitem{etcsd}S.\ Dimopoulos and L.\ Susskind, \Journal{\NPB}{155}{237}
{1979}.
%
\bibitem{etceekl}E.~Eichten and K.~Lane, \Journal{\PLB}{90}{125}{1980}.
%
\bibitem{ichep}For a review of technicolor and its signatures up to 1996,
see K.~Lane, {\it Non-Supersymmetric Extensions of the Standard Model},
hep-ph/9610463, plenary talk at the 28th International Conference on High
Energy Physics, edited by Z.~Ajduk and A.~K.~Wroblewski, Vol.~I, p.~367,
Warsaw, July~25-31, 1996.
%
\bibitem{wtc}B.~Holdom, \Journal{\PRD}{24}{1441}{1981};
\Journal{\PLB}{150}{301}{1985}; T.~Appelquist, D.~Karabali and L.~C.~R.
Wijewardhana, \Journal{\PRL}{57}{957}{1986}; T.~Appelquist and
L.~C.~R.~Wijewardhana, \Journal{\PRD}{36}{568}{1987}; K.~Yamawaki, M.~Bando
and K.~Matumoto, \Journal{\PRL}{56}{1335}{1986}; T.~Akiba and T.~Yanagida,
\Journal{\PLB}{169}{432}{1986}.
%
\bibitem{multi}K. Lane and E. Eichten,
\Journal{\PLB}{222}{274}{1989}; K.~Lane and M.~V.~Ramana,
\Journal{\PRD}{44}{2678}{1991}.
%
\bibitem{balaji}B.~Balaji, \Journal{\PRD}{53}{1699}{1996}.
%
\bibitem{glasgow}K.~Lane, {\em Technicolor and Precision Tests of the
Electroweak Interactions}, Proceedings of the 27th International Conference
on High Energy Physics, edited by P.~J.~Bussey and I.~G.~Knowles, Vol.~II,
p.~543, Glasgow, June 20--27, 1994.
%
\bibitem{pettests}B.~W.~Lynn,
M.~E.~Peskin and R.~G.~Stuart, in {\em Trieste Electroweak 1985}, 213 
(1985); M.~E.~Peskin and T.~Takeuchi, \Journal{\PRL}{65}{964}{1990};
A.~Longhitano, \Journal{\PRD}{22}{1166}{1980};
\Journal{\NPB}{188}{118}{1981}; R.~Renken and M.~Peskin,
\Journal{\NPB}{211}{93}{1983}; M.~Golden and L.~Randall,
\Journal{\NPB}{361}{3}{1990}; B.~Holdom and J.~Terning,
\Journal{\PLB}{247}{88}{1990}; A.~Dobado, D.~Espriu and M~J.~Herrero,
\Journal{\PLB}{255}{405}{1990}; H.~Georgi, \Journal{\NPB}{363}{301}{1991}.
%
\bibitem{toprefs}F.\ Abe, et al., The
CDF Collaboration, \Journal{\PRL} {73}{225}{1994};
\Journal{\PRD}{50}{2966}{1994}; \Journal{\PRL}{74}{2626} {1995}; S.~Abachi,
et al., The D\O\ Collaboration, \Journal{\PRL}{74}{2632}{1995}.
%
%
\bibitem{zbbth}R.~S.~Chivukula, S.~B.~Selipsky, and E.~H.~Simmons,
\Journal{\PRL}{69}{575}{1992}; R.~S.~Chivukula, E.~H.~Simmons, and
J.~Terning, \Journal{\PLB}{331}{383}{1994}, and references therein.
%
\bibitem{tctwohill}C.~T.~Hill, \Journal{\PLB}{345}{483}{1995}.
%
\bibitem{topcondref}Y.~Nambu, in {\it New Theories in Physics}, Proceedings
of the XI International Symposium on Elementary Particle Physics,
Kazimierz, Poland, 1988, edited by Z.~Adjuk, S.~Pokorski and A.~Trautmann
(World Scientific, Singapore, 1989); Enrico Fermi Institute Report
EFI~89-08 (unpublished); V.~A.~Miransky, M.~Tanabashi and K.~Yamawaki,
\Journal{\PLB}{221}{171}{1989}; {\em Mod.~Phys.~Lett.}~{\bf A4}, 1043
(1989); W.~A.~Bardeen, C.~T.~Hill and M.~Lindner,
\Journal{\PRD}{D41}{1647}{1990}.
\bibitem{topcref}C.~T. Hill, \Journal{\PLB}{266}{419}{1991}; S.~P.~Martin,
\Journal{\PRD}{45}{4283}{1992}; {\it ibid}~{\bf D46}, 2197 (1992);
\Journal{\NPB}{398}{359}{1993}; M.~Lindner and D.~Ross,
\Journal{\NPB}{B370}{30}{1992}; R.~B\"{o}nisch,
\Journal{\PLB}{268}{394}{1991}; C.~T.~Hill, D.~Kennedy, T.~Onogi, H.~L.~Yu,
\Journal{\PRD}{47}{2940} {1993}.
%
\bibitem{tctwoklee}K.~Lane and E.~Eichten, \Journal{\PLB}{352}{382}{1995};
K.~Lane, \Journal{\PRD}{54}{2204}{1996}.
%
%
\bibitem{ellis}J.~Ellis, M.~K.~Gaillard, D.~V.~Nanopoulos and P.~Sikivie,
\Journal{NPB}{529}{1981}.
%
\bibitem{tpitev}E.~Eichten and K.~Lane, \Journal{\PLB}{388}{803}{1996}.
%
\bibitem{elwtev}E.~Eichten, K.~Lane and J.~Womersley,
\Journal{PLB}{405}{305}{1997}.
%
\bibitem{elwfmc}E.~Eichten, K.~Lane and J.~Womersley, ``Narrow Technihadron
Production at the First Muon Collider'', in preparation. Also see
J.~Womersley, ``Technihadron Production at a Muon Collider'', to appear in
the FMC proceedings.
%
\bibitem{bdcth}D.~Bobrescu and C.~T.~Hill, 
FERMILAB-PUB-97-409-T, hep-ph/9712319 (Dec.~1997). 

\end{thebibliography}
\end{document}